\documentclass[aps,prl,amsmath,amssymb,superscriptaddress]{revtex4}
\usepackage{graphicx}
\usepackage{amssymb}
\usepackage{natbib}
\usepackage[compact]{titlesec}
\usepackage[rawfloats=true]{floatrow}
\usepackage{array}
\usepackage{chngcntr}
\usepackage{color}

\restylefloat{figure}

\begin{document}

\title{Anisotropic spin fluctuations in detwinned FeSe}
\author{Tong Chen}
\affiliation{Department of Physics and Astronomy, Rice University, Houston, Texas 77005, USA}
\author{Youzhe Chen}
\affiliation{Department of Physics and Astronomy, Johns Hopkins University, Baltimore, Maryland 21218, USA}
\author{Andreas Kreisel}
\affiliation{Institut f\"{u}r Theoretische Physik, Universit\"{a}t Leipzig, D-04103 Leipzig, Germany}
\author{Xingye Lu}
\email{luxy@bnu.edu.cn}
\affiliation{Center for Advanced Quantum Studies and Department of Physics, Beijing Normal University, Beijing 100875, China}
\author{Astrid Schneidewind}
\affiliation{J\"{u}lich Center for Neutron Sciences, Forschungszentrum J\"{u}lich GmbH, Outstation at MLZ, D-85747 Garching, Germany}
\author{Yiming Qiu}
\affiliation{NIST Center for Neutron Research, National Institute of Standards and Technology, Gaithersburg, Maryland 20899, USA}
\author{Jitae Park}
\affiliation{Heinz Maier-Leibnitz Zentrum (MLZ), Technische Universit\"{a}t M\"{u}nchen, D-85747 Garching, Germany}
\author{Toby G. Perring}
\affiliation{ISIS Facility, STFC Rutherford-Appleton Laboratory, Didcot OX11 0OX, UK}
\author{J Ross Stewart}
\affiliation{ISIS Facility, STFC Rutherford-Appleton Laboratory, Didcot OX11 0OX, UK}
\author{Huibo Cao}
\affiliation{Neutron Scattering Division, Oak Ridge National Laboratory, Oak Ridge, Tennessee 37831, USA}
\author{Rui Zhang}
\affiliation{Department of Physics and Astronomy, Rice University, Houston, Texas 77005, USA}
\author{Yu Li}
\affiliation{Department of Physics and Astronomy, Rice University, Houston, Texas 77005, USA}
\author{Yan Rong}
\affiliation{Center for Advanced Quantum Studies and Department of Physics, Beijing Normal University, Beijing 100875, China}
\author{Yuan Wei}
\affiliation{Institute of Physics, Chinese Academy of Sciences, Beijing 100190, China}
\author{Brian M. Andersen}
\affiliation{Niels Bohr Institute, University of Copenhagen, Lyngbyvej 2, DK-2100 Copenhagen, Denmark}
\author{P. J. Hirschfeld}
\affiliation{Department of Physics, University of Florida, Gainesville, Florida 32611, USA}
\author{Collin Broholm}
\affiliation{Department of Physics and Astronomy, Johns Hopkins University, Baltimore, Maryland 21218, USA}
\affiliation{NIST Center for Neutron Research, National Institute of Standards and Technology, Gaithersburg, Maryland 20899, USA}
\author{Pengcheng Dai}
\email{pdai@rice.edu}
\affiliation{Department of Physics and Astronomy, Rice University, Houston, Texas 77005, USA}
\affiliation{Center for Advanced Quantum Studies and Department of Physics, Beijing Normal University, Beijing 100875, China}

\begin{abstract}
Superconductivity in FeSe emerges from a nematic phase that breaks four-fold rotational symmetry in the iron plane. This phase may arise from orbital ordering, spin fluctuations, or hidden magnetic quadrupolar order. Here we use inelastic neutron scattering on a mosaic of single crystals of FeSe detwinned by mounting on a BaFe$_2$As$_2$ substrate to demonstrate that spin excitations are most intense at the antiferromagnetic wave vectors ${\bf Q}_{\rm AF}=(\pm 1,0)$ at low energies $E = 6$−$11$ meV in the normal state. This two-fold ($C_2$) anisotropy is reduced at lower energies 3-5 meV, indicating a gapped four-fold ($C_4$) mode. In the superconducting state, however, the strong nematic anisotropy is again reflected in the spin resonance ($E = 3.7$ meV) at ${\bf Q}_{\rm AF}$ with incommensurate scattering around 5-6 meV. Our results highlight the extreme electronic anisotropy of the nematic phase of FeSe and are consistent with a highly anisotropic superconducting gap driven by spin fluctuations. 
\end{abstract}

\maketitle

High-transition temperature 
superconductivity in copper and iron based materials emerges from their 
antiferromagnetic (AF) ordered nonsuperconducting   
parent compounds \cite{scalapino}. 
While the parents of copper oxide superconductors are Mott insulators with a simple checkerboard AF structure \cite{scalapino}, most iron pnictide
parent materials exhibit a tetragonal-to-orthorhombic structural transition 
at $T_s$ ($<295$ K) and form twin-domains before ordering antiferromagnetically at $T_N$ ($T_s\geq T_N$) \cite{dai}. Therefore, one must detwin iron pnictides in order to measure their intrinsic electronic properties below $T_s$. By applying a uniaxial pressure along one-axis of the orthorhombic lattice to detwin the sample, an in-plane resistivity anisotropy
has been observed in strained iron pnictides BaFe$_{2-x}T_x$As$_2$ (where $T$
is Co or Ni) above $T_s$ \cite{JHChu2010,HHKuo2016}.
The resistivity anisotropy has been ascribed to
an electronic nematic phase that spontaneously breaks the rotational symmetry while preserving the translation symmetry of
the underlying lattice and is established in the temperature regime
below $T_s$ and above $T_N$ \cite{Fernandes14,AEBohmerCRP}.
Below $T_N$, the AF structure is collinear, consisting of columns of
antiparallel spins along the orthorhombic $a_o$ axis and parallel
spins along the $b_o$ axis with an in-plane AF ordering wave vector
${\bf Q}_{\rm AF}=(\pm 1,0)$ in reciprocal space \cite{dai}.

The highly unusual iron-based superconductor 
FeSe exhibits an orthorhombic structural distortion and superconductivity without static AF order [Fig. 1(a)] \cite{Hsu,McQueen09,bohmer}. Although the nematic phase in FeSe is established below $T_s$ ($\approx 90$ K) \cite{McQueen09}, it has been argued that nematic order and superconductivity are 
induced by orbital fluctuations [Fig. 1(b)] \cite{CCLee2009,Baek15,anna15,Yamakawa16,Onari16}, forming a sign-preserving $s^{++}$-wave electron pairing and therefore would be fundamentally different from
other iron-based superconductors \cite{Kontani10}. Alternatively, the absence of
static AF order in FeSe has been interpreted as evidence for a quantum
paramagnet arising from the $d$-orbital spin-1 localized
iron moments \cite{FWang15,Glasbrenner15}.  Here, the nematic phase is driven by
magnetic frustration due to competition between low-energy spin fluctuations associated
with AF collinear order and those associated with
various types of staggered order \cite{Qwang16}.
Third, the nematic superconductivity in FeSe without AF order may arise from a frustration-induced nematic quantum spin liquid state with melted AF order \cite{She17}. This model predicted 
a dramatic suppression of the magnetic spectral weight at ${\bf Q}=(0,\pm 1)$ in a detwinned sample, and
explained the observed superconducting gap anisotropy by 
angle resolved photo-emission spectroscopy (ARPES) \cite{MYi17,Coldea18,Liu2018} 
and scanning tunneling microscopy (STM) \cite{CLSong17,sprau17,Kostin2018} experiments by an 
orbital dependent Hund's coupling \cite{She17}. 
Forth, the nematic order may arise from a hidden magnetic quadrupolar order \cite{ZWang16,Lai17}.
Finally, the  nematic phase and superconductivity in FeSe has also been described by
itinerant electrons interacting among quasi-nested hole-electron Fermi surfaces \cite{Mukherjee15,Kreisel15}, as in other iron based
superconductors \cite{Hirschfeld16}.
In this picture, the electronic correlation effect is taken into account
by orbital-dependent quasiparticle weights \cite{sprau17,Kreisel18}. Without electron correlation effects, spin fluctuations
in the nematic phase below $T_s$ exhibit only a minor $C_4$ asymmetry. Including correlations in the theoretical calculations render the spin fluctuations highly $C_2$ symmetric with negligible weight at $(0,\pm 1)$, and a neutron spin resonance exhibited only at ${\bf Q}_{\rm AF}=(\pm 1,0)$ driven by the $d_{yz}$ orbitals \cite{Kreisel18}.
Approaches based on localized
models with magnetic quadrupolar order have also predicted a strong suppression of low-energy $(0,\pm 1)$ 
intensity \cite{Lai17}.

In recent inelastic neutron scattering (INS) experiments on twinned FeSe \cite{Qwang16,wang16,MWMa17},
well-defined low-energy ($E<15$ meV) spin fluctuations are found at ${\bf Q}_{\rm AF}=(\pm 1,0)$
and its twin-domain positions $(0,\pm 1)$ in the nematic phase below $T_s$.
On cooling below $T_c$, a neutron spin resonance, a key signature of unconventional
superconductivity \cite{scalapino}, appears at $E\approx 4$ meV and sharply
peaks at the $(\pm 1,0)$ and $(0,\pm 1)$ positions \cite{wang16,MWMa17}. 
Figure 1(c) shows the energy dependence of the magnetic scattering $S(E)$ integrated 
around ${\bf Q}_{\rm AF}$ obtained from our high-resolution INS experiments (see Methods).
In the normal state, the magnetic scattering is gapless above $E=0.5$ meV and increases in intensity
with increasing energy [Fig. 2(a)].  In addition to having a weak peak
around $E\approx 3.2$ meV, we find that the scattering changes from well-defined
commensurate peaks centered around ${\bf Q}_{\rm AF}$ below $E=3.625\pm 0.125$ meV [Figs. 2(b), 2(c)] to a peak with flattish top at $E=5.625\pm 0.125$ meV [Fig. 2(d)]. Upon cooling to below $T_c$ in 
the superconducting state, the spin excitation spectra open a gap below $E\approx 2.5$ meV
[Figs. 2(e), 2(f)], form a commensurate resonance
at $E=3.6$ meV [Fig. 2(g)], and exhibit ring-like incommensurate 
scattering at $E=5.25\pm 0.075$ meV [Fig. 2(f)]. The dispersive ring-like incommensurate resonance is also seen in hole-doped 
Ba$_{0.67}$K$_{0.33}$(Fe$_{1-x}$Co$_x$)$_2$As$_2$ superconductors \cite{RZhang2018}.

Although these results on twinned FeSe
suggest that spin fluctuations play an important role in the superconductivity of FeSe,
they provide no information on the possible orbital selective
nature of the fluctuations that may lead to a highly anisotropic
 electron pairing state \cite{Kreisel18,Nica17,kreisel17,Benfatto18,Kang2018,She17}.
From STM quasiparticle interference measurements on a single domain (detwinned) FeSe, where the Fermi
surface geometry of electronic bands can be
determined in the nematic phase, sign-reversed superconducting gaps are found
at the hole [$\Gamma$ or ${\bf Q}=(0,0)$] and electron [$X$ or ${\bf Q}_{\rm AF}=(1,0)$] Fermi surface states
derived from $d_{yz}$ orbitals of the Fe
atoms along the orthorhombic $a_o$-axis direction [Figs. 1(a) and 1(b)] \cite{sprau17}.
Moreover, similar STM measurements show that the same orbital selective self-energy effects are present already in the normal state of FeSe above $T_c$ \cite{Kostin2018}.

\begin{figure}[t]
\includegraphics[scale=.70]{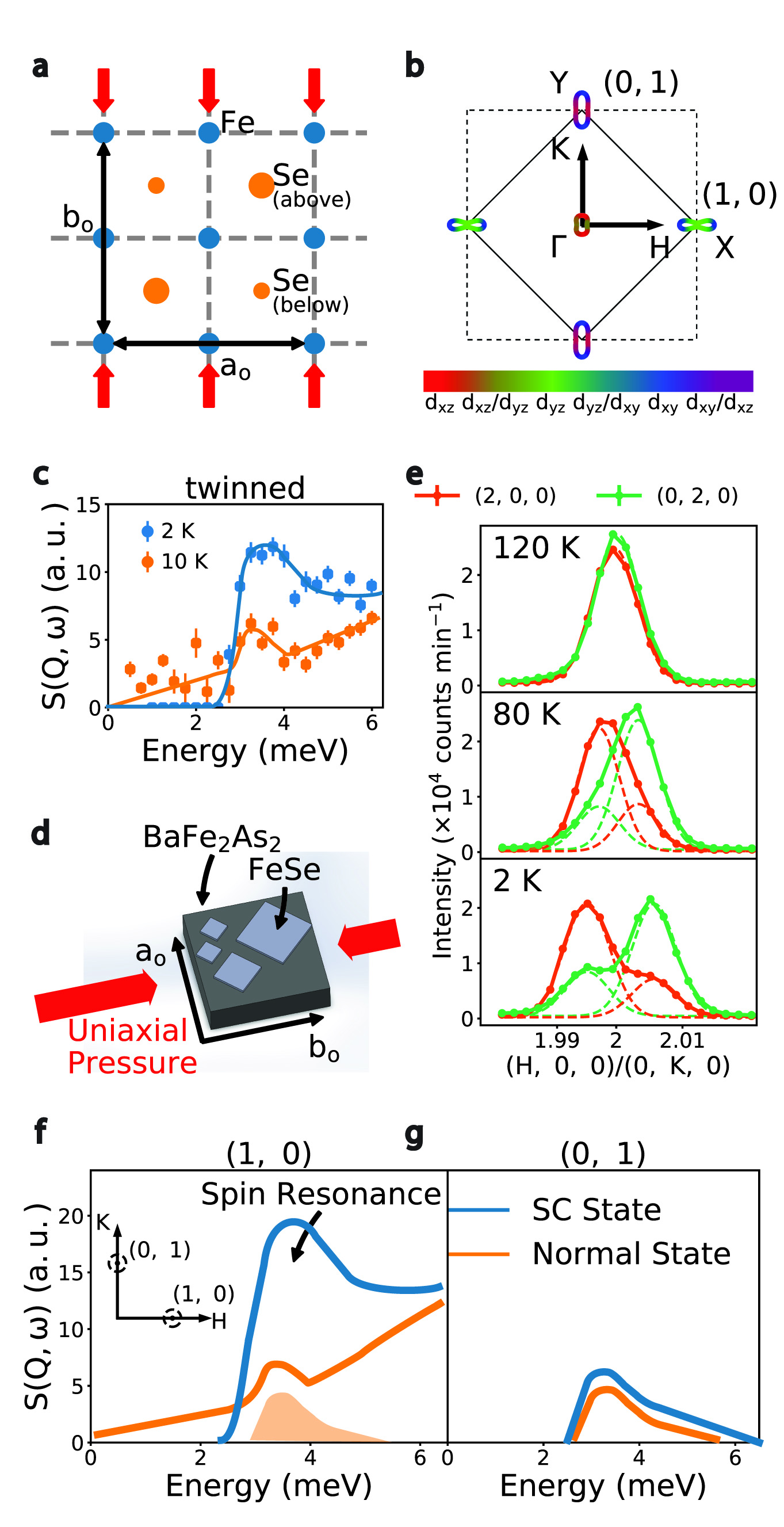}
\caption{{\bf Crystal structure, Fermi surface and neutron scattering of FeSe}
(a) Crystal structure of FeSe, where blue and orange colors mark Fe and Se positions, respectively.
The red arrows indicate the uniaxial strain direction applied through detwinned BaFe$_2$As$_2$. $a_o$
and $b_o$ are the orthorhombic lattice parameters (double-sided black arrows) in the nematic phase. Gray dashed lines are guides for the eye.
(b) Hole/electron Fermi surfaces of the tight binding model for FeSe \cite{sprau17}. Color
indicates orbital character of the Fermi surfaces, where red, green, and blue indicate
$d_{xz}$, $d_{yz}$, and $d_{xy}$ orbitals of the Fe atom.  Fermi surface nesting of
$\Gamma\rightarrow X$ and $\Gamma\rightarrow Y$ corresponds to $(1,0)$ and $(0,1)$
in reciprocal lattice units (r.l.u.), 
where $(H,K)=(q_xa_o/2{\pi},q_yb_o/2{\pi})$ are in-plane 
Miller indices of the orthorhombic lattice, respectively (see Methods).
(c) Energy dependence of the measured magnetic scattering ($S(Q,E)$) integrated around $(1,0)$ above ($T=10$ K) 
and below ($T=2$ K) $T_c$ ($\approx 8$ K) on twinned FeSe (Supplemental Fig. S5). 
The vertical error bars indicate the statistical errors of one standard deviation. 
(d) Schematic diagram of the sample arrangement, FeSe samples are glued on large single crystals of
BaFe$_2$As$_2$ under uniaxial pressure of $\sim$20 MPa \cite{Lu14,Lu18}, 
where red arrows indicate pressure direction.
(e) Wave vector scans of nuclear $(2,0,0)$ and $(0,2,0)$ Bragg peaks of FeSe on an assembly of BaFe$_2$As$_2$ single crystals at different temperatures. The dashed lines indicate the single Gaussian components of the fitting.
(f,g) Schematic illustration of the magnetic scattering
at $(1,0)$ and $(0,1)$ in the normal and superconducting (SC) states estimated from the twinned and detwinned samples (Supplemental Fig. S10). Shaded region in (f) is $(0,1)$ data from (g).\label{Fig1}}
\end{figure}

If superconductivity in FeSe arises from the quasiparticle excitations between hole and electron pockets [Fig. 1(b)] that are indeed orbital selective \cite{sprau17,Kostin2018}, detwinned crystals should exhibit a strong anisotropy of the low energy spin excitations.  In particular, it is expected that the neutron spin resonance associated with $s^{\pm}$ superconductivity \cite{wang16,MWMa17} should only occur
along the orthorhombic $a_o$-axis direction at ${\bf Q}_{\rm AF} = (\pm 1,0)$ in a detwinned FeSe,
as the orbital selective superconducting gap with the $d_{yz}$ orbital character is large for scattering vectors along the
$a_o$-axis \cite{sprau17}. Similarly, orbital dependent Hund's coupling in a nematic quantum spin
liquid of FeSe can also induce a large superconducting gap and spin excitation anisotropy \cite{She17}.  
To test these hypothesis,
we used INS to study the low-energy spin
fluctuations in detwinned FeSe [Figs. 1(d)-1(e)].
In the normal state, spin fluctuations from 6-11 meV are centered around ${\bf Q}_{\rm AF}$
with negligible intensity at $(0,\pm 1)$, thus exhibiting a pronounced $C_2$ rotational symmetry as predicted by these theoretical approaches \cite{She17,Lai17,Kreisel18}. 
By contrast, for energies between 3-5 meV, the spin fluctuations have a  
$C_4$ rotational symmetry magnetic component as shown in the schematic illustration in Figs. 1(f) and 1(g) which is based on combining experimental evidences from multiple instruments (Supplementary Fig. S10), possibly corresponding to a localized mode in both wave vector and energy.  On cooling below $T_c$, the resonance only appears
at ${\bf Q}_{\rm AF} = (1,0)$ [Figs. 1(f) and 1(g)], consistent with the STM observation that
superconducting gaps are extremely anisotropic 
with minima at the tips of the elliptical pockets.
Therefore, while the normal state $C_4$ rotational symmetry magnetic component in the 3-5meV range is not anticipated, 
the anisotropic superconductivity-induced resonance is consistent with theoretical expectations \cite{She17,sprau17}.

\begin{figure}[t]
\includegraphics[scale=.50]{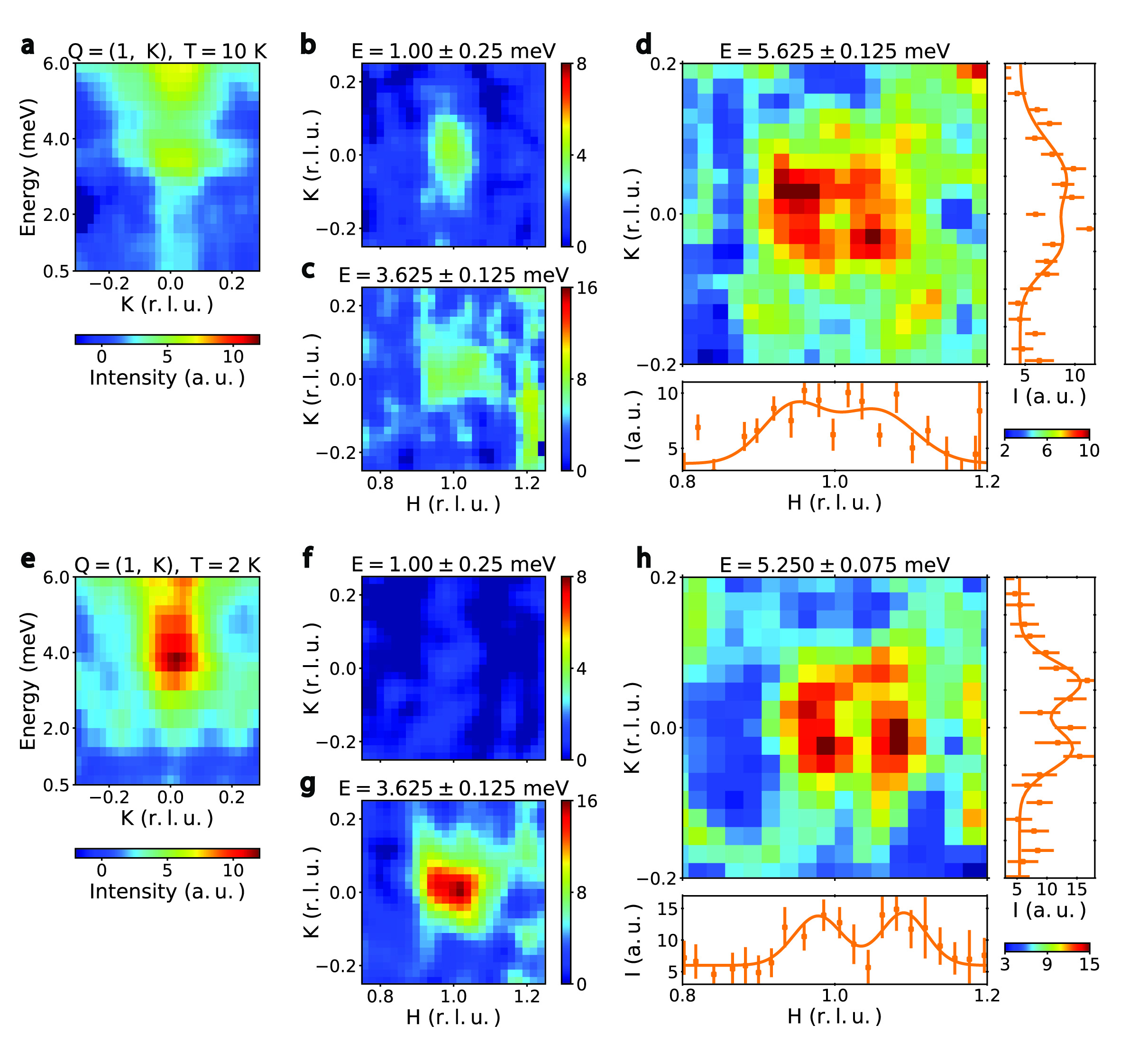}
\caption{{\bf Low-energy spin fluactuations in twinned FeSe below and above $T_c$.}
Two-dimensional images of wave vector and energy dependence of 
spin fluctuations at (a) $T=10$ K, (e) $T=2$ K.
Wave vector dependence of spin fluctuations in the $(H,K)$ plane 
at energies (b,f) $E=1\pm 0.25$ meV, (c,g) $E=3.6\pm 0.125$ meV, 
(d) $E=5.62\pm 0.125$ meV.  Cuts along the $[H,0]$ and $[1,K]$ 
directions with a width of $\pm 0.04$ r.l.u. show flattish top scattering near $(1,0)$.
(h) $E=5.25\pm 0.075$ meV. Incommensurate scattering is clearly seen
through the identical cuts along the $[H,0]$ and $[1,K]$ directions. This feature is missed in previous work \cite{Qwang16,wang16,MWMa17} due to the small incommensurability and narrow
energy range.
Panels (a,b,c,d) and (e,f,g,h) are at $T=10$ K and $T=2$ K, respectively.
Solid lines are fitting with the sum of two Gaussians to the data. The vertical error bars indicate the statistical errors of one standard deviation.
}
\end{figure}

\begin{figure}[t]
\includegraphics[scale=.80]{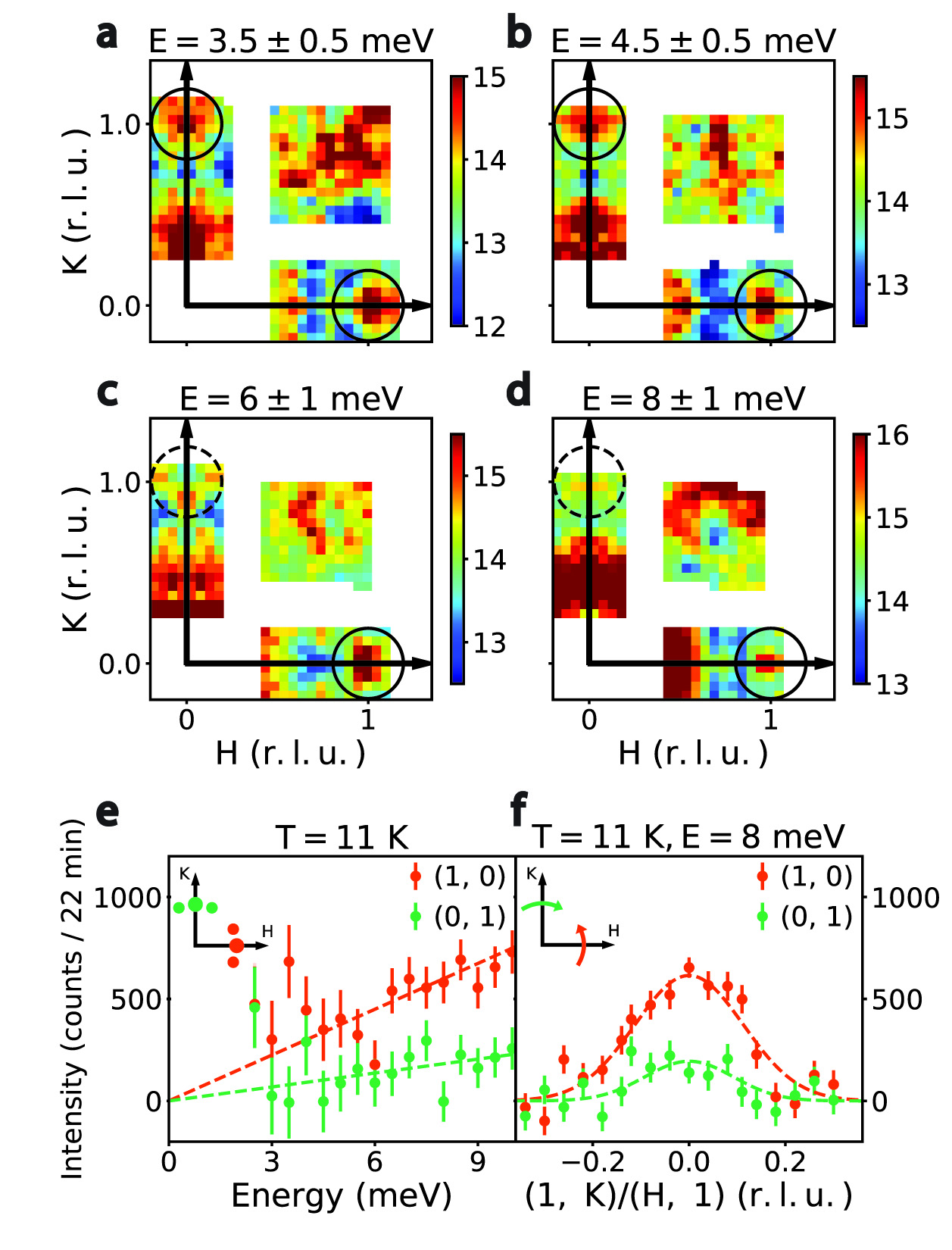}
\caption{{\bf Normal state spin fluctuations in detwinned FeSe}.
The two-dimensional images of spin fluctuations at (a) $E=3.5\pm 0.5$ meV, (b)
$E=4.5\pm 0.5$ meV, (c) $E=6\pm 1$ meV, and $E=8.5\pm 1$ meV.  The data are
collected at $T=12$ K using MAPS chopper spectrometer with incident neutron 
energy of $E_i=38$ meV along the
$c$ axis and are folded to improve statistics. The scattering near wave vector 
$(1,1)$ is background and not magnetic in origin (Supplementary Figs. S6-S8).  
(e) Energy dependence of the scattering $(1,0)$ and $(0,1)$ above background
at $T=11$ K.
The positions of signal and background are marked as large and small spots in the inset. (f) Wave vector
scans at $E=8$ meV along the $[1,K]$ (green) and $[H,1]$ direction at $T=11$ K.  Linear backgrounds
have been subtracted from the data. The vertical error bars indicate the statistical errors of one standard deviation. \label{Fig3}
}

\end{figure}

To detect anisotropic spin fluctuations by INS \cite{wang16,MWMa17}, one needs to co-align hundreds of single crystal
FeSe samples. These are grown by chemical vapor transport method and
are about 1-3 mm$^2$ in size while few $\mu$m in thickness (see Methods) \cite{bohmer}. Therefore, the 
most difficult part
of carrying out INS experiments on FeSe is to simultaneously detwin
hundreds of samples. In previous work on iron pnictides,
we were able to completely detwin large (on the order of 0.5-1 cm$^2$ by few mm in thickness)
single crystals of BaFe$_2$As$_2$ using a mechanical
uniaxial pressure device \cite{Lu14,Lu18}. By gluing many oriented FeSe on uniaxial
pressured BaFe$_2$As$_2$ shown
schematically in Fig. 1(d), we were able to simultaneously detwin many
FeSe single crystals required for INS experiments (Supplementary Fig. S2).
Figure 1(e) shows the temperature dependence of rocking scans along the
$[H,0,0]$ and $[0,K,0]$ directions on multiple FeSe on BaFe$_2$As$_2$
assemblies. Below $T_s\approx 90$ K, we see a clear splitting of the lattice constants.
By comparing the scattering intensity of the $(2,0,0)$ and $(0,2,0)$ nuclear Bragg peaks, we find that the FeSe sample
assembly has detwinning ratio of 
$\eta=[I(2,0,0)_o-I(0,2,0)_o]/[I(2,0,0)_o+I(0,2,0)_o]\approx 50\%$ at 2 K [Fig. 1(e)],
where $I(2,0,0)_o$ and $I(0,2,0)_o$ are the observed Bragg peak intensity at $(2,0,0)$ and $(0,2,0)$, respectively, below $T_s$ (Supplementary Figs. S3 and S4).

\begin{figure}[t]
\includegraphics[scale=.8]{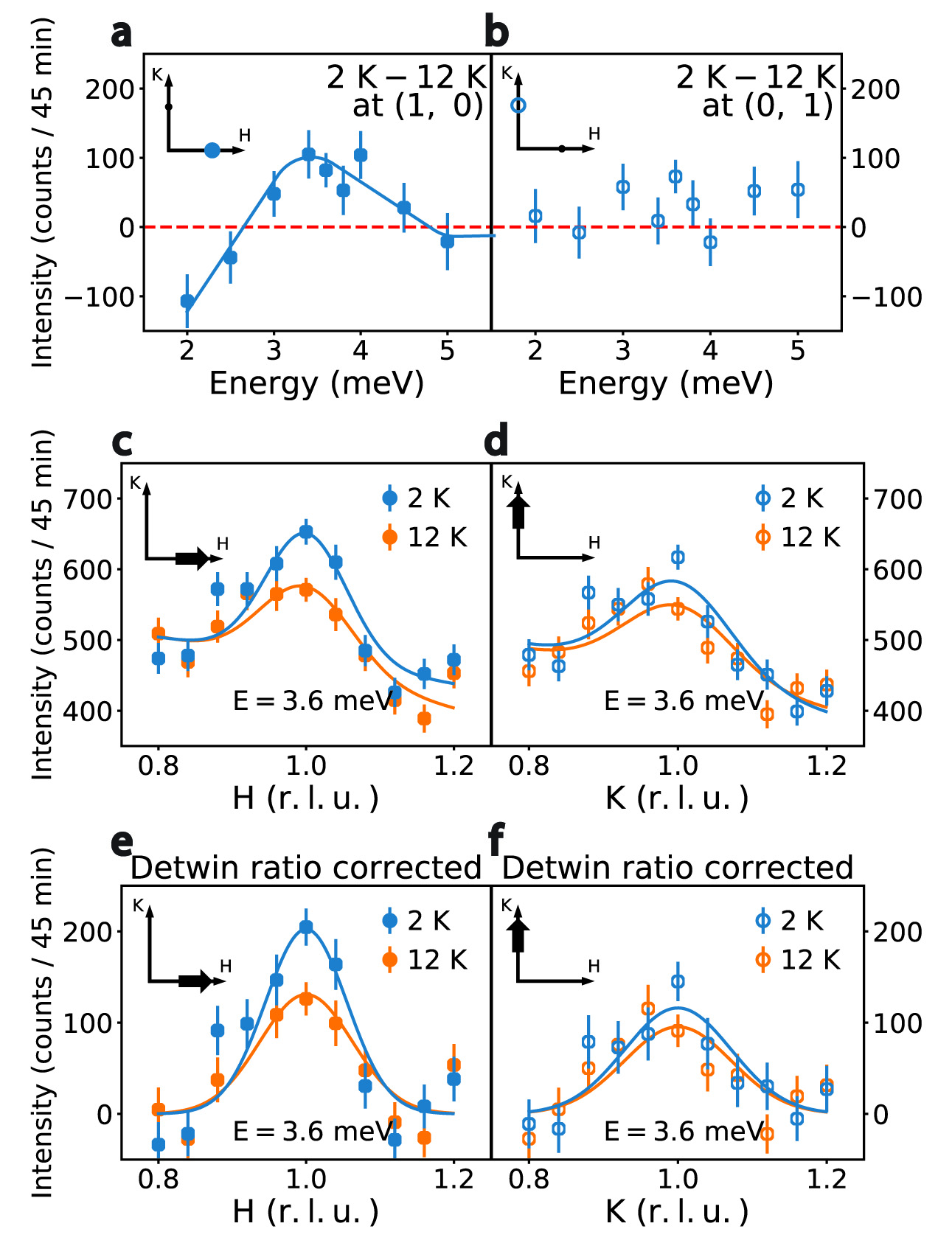}
\caption{{\bf Effect of superconductivity on low-energy spin fluctuations of detwinned FeSe.}
Difference of the scattering in the superconducting state (below $T_c$) and the normal state plotted as function of energy for 
the momentum transfer 
(a) $(1,0)$ and
(b) $(0,1)$.  The peak seen at $E\approx 3.6$ meV in (a) marks the neutron spin resonance. The solid blue line and dashed red lines are guides to the eye.
(c) Wave vector scans below and above $T_c$ at $E=3.6$ meV and $(1,0)$. (d) Similar scans
at $(0,1)$. (e) $(1,0)$ and (f) $(0,1)$ scans with background subtracted and detwinning ratio corrected (Supplemental Fig. S9). The solid lines in
(c,d) and (e,f) are Gaussian fits to the data before and after linear background subtraction.
The vertical error bars indicate the statistical errors of one standard deviation.
}
\end{figure}

In order to understand the effect of detwinning FeSe, we first need to determine the wave vector 
and energy dependence of the magnetic scattering $S({\bf Q},E)$ in twinned samples (See Methods and supplementary Fig. S5). Figures 2(a)
and 2(e) show the energy dependence of the magnetic scattering along the $[1,K,0]$ direction
above and below $T_c$, respectively. In the normal state at $T=10$ K, the scattering
is gapless above $E=0.5$ meV and exhibits a weak peak around $E=3.2$ meV [Fig. 2(a)].
The spin excitations are centered around ${\bf Q}_{\rm AF} = (1,0)$ at $E=1\pm 0.25$ meV 
[Fig. 2(b)] and $3.625\pm 0.125$ meV [Fig. 1(c)]. At $E=5.625\pm 0.125$ meV, the spin excitations
have a flattish top as revealed by wave vector cuts along the $[H,0]$ and $[1,K]$ directions [Fig. 1(d)]. In the superconducting
state at $T=2$ K, a superconductivity-induced spin gap opens below $E\approx 2.5$ meV and
a resonance forms around $E=3.6$ meV [Figs. 1(c), 2(e)].
 This is confirmed by the vanishing signal at $E=1\pm 0.25$ meV [Fig. 2(f)] and enhanced
magnetic scattering at $3.625\pm 0.125$ meV [Fig. 2(g)]. In addition, the resonance  
is clearly centered at the commensurate ${\bf Q}_{\rm AF} = (1,0)$ position [Fig. 2(g)].  
However, on increasing energy to $E=5.25\pm0.075$ meV, we see clear incommensurate ring-like 
magnetic scattering centered around ${\bf Q}_{\rm AF} = (1,0)$, as confirmed by
wave vector cuts along the $[H,0]$ and $[1,K]$ directions [Fig. 2(h)]. The incommensurate
scattering intensity in the superconducting state is higher than that of the normal state, suggesting it is a part of the dispersive resonance.  
In previous work, a dispersive ring-like neutron
spin resonance has been seen in the hole-doped 
BaFe$_2$As$_2$ family of materials,
where the incommensurate scattering has been ascribed to quasiparticle excitations from mismatched hole and electron Fermi surfaces \cite{RZhang2018}.

Figure 3 summarizes the energy evolution of the normal state spin fluctuations at ${\bf Q}_{\rm AF} = (1,0)$
and $(0,1)$ in the $(H,K)$ plane in partially detwinned FeSe. Since our FeSe single crystals are mounted on surfaces of BaFe$_2$As$_2$, one should also see spin fluctuations from BaFe$_2$As$_2$ at approximately the same position in reciprocal space. However, the spin waves in BaFe$_2$As$_2$ are gapped below $\sim$10 meV
in the low-temperature AF ordered state \cite{Matan09,CWang13}, meaning that spin fluctuations
at ${\bf Q}_{\rm AF} \approx (1,0)$ and $(0,1)$ below 10 meV must originate from FeSe.
Figures 3(a) and 3(b) show constant-energy cuts in the $(H,K)$ plane for energy transfers of
$E=3.5\pm 0.5$ and $4.5\pm 0.5$ meV, respectively, in the normal
state at $T=12$ K.  We see clear evidence for magnetic scattering at
${\bf Q}_{\rm AF} \approx (1,0)$ and $(0,1)$ with about the same strength (Supplementary Figs. S6a-S6d), suggesting
a possible mode that has $C_4$ rotational symmetry in the normal state.
On increasing energies to $E=6\pm 1$ and $8\pm 1$ meV, the scattering at
${\bf Q}_{\rm AF} \approx (1,0)$ becomes much stronger than those at $(0,1)$, suggesting
that spin fluctuations become highly $C_2$ symmetric at these energies [Figs. 3(c) and 3(d)].
To confirm these results, we carried out energy scans at
${\bf Q}_{\rm AF} \approx (1,0)$ and $(0,1)$ from 2.5 meV to 11 meV as shown in Fig. 3(e) (Supplementary Fig. S6e).
 From 6 meV to 11 meV, magnetic scattering 
at $(1,0)$ increase in intensity with increasing energy approximately two times faster than the increase of magnetic scattering at $(0,1)$. Figure 3(f) shows
wave vector scans approximately along the $[1,K]$ and $[H,1]$ directions at $E=8$ meV (see $E=3.6$ meV data in supplementary Fig. S6f). The scattering intensity at $(1,0)$ dominated the signal while spin fluctuations at  
$(0,1)$ are only 1/3 of that at $(1,0)$. After taking into account the finite detwinning ratio $\eta$ of the FeSe samples (see supplementary information), there is almost  
no magnetic scattering at $(0,1)$ above the background. 
These results are consistent with Figs. 3(c) and 3(d), suggesting that the spin fluctuations between 6-10 meV are strongly $C_2$ symmetric.

\begin{figure}[t]
\includegraphics[scale=.8]{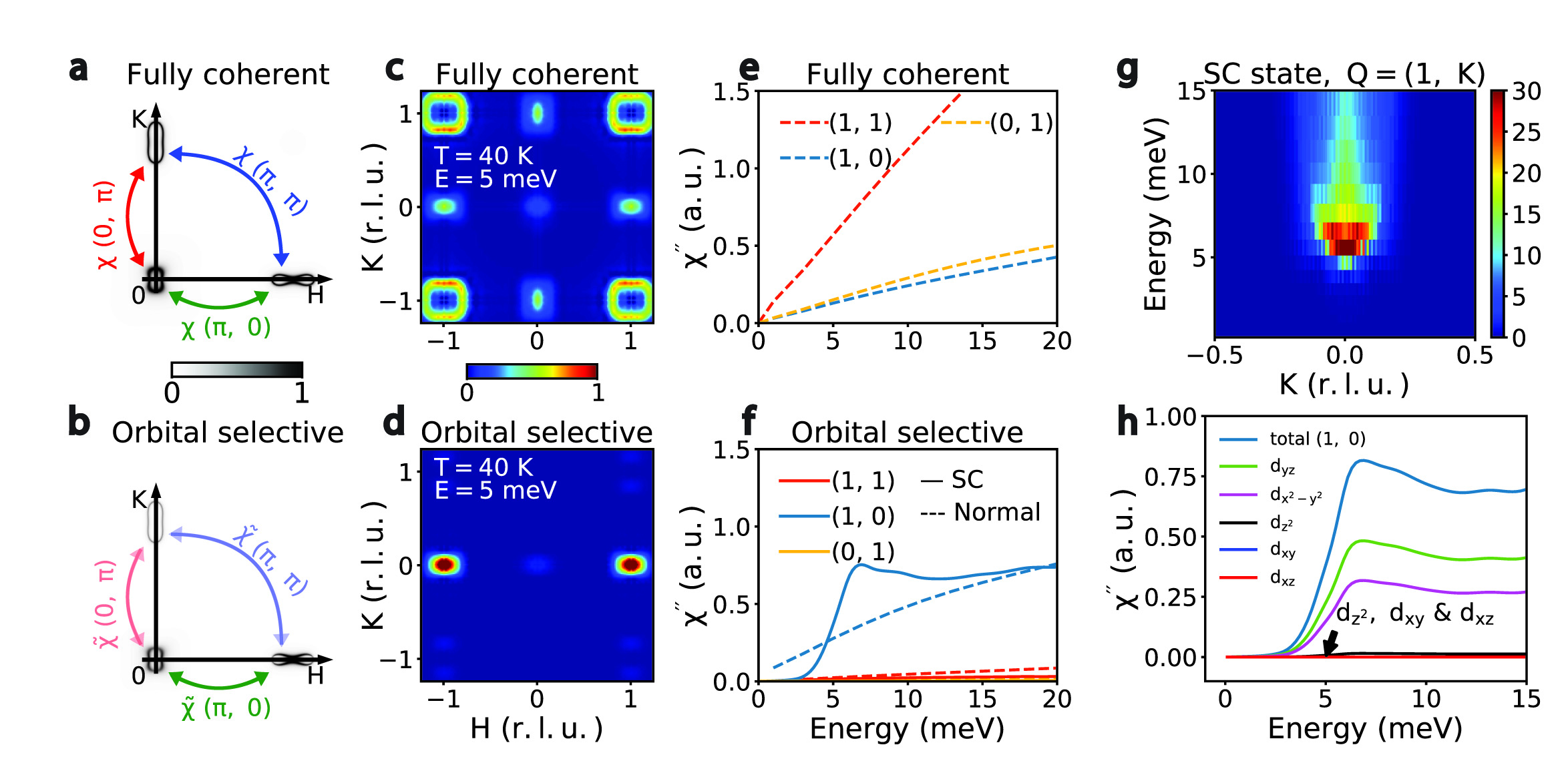}
\caption{{\bf Theoretical calculations of the spin fluctuations in detwinned FeSe.}
Map of the spectral function at zero energy for our model of the electronic structure of FeSe in (a) the fully coherent case with quasiparticle weights ($Z_{\ell}=1$) and (b) the orbital-selective case where
reduced quasiparticle weights ($Z_{\ell}<1$) weaken the spin-fluctuations at certain momentum transfer (blue and red arrows) while the spin fluctuations stemming mostly from the $d_{yz}$ orbital become dominant (green arrow).
(c) For the fully coherent model, one obtains in the normal state large contributions to the dynamical structure factor
close to $(\pm 1, \pm 1)$ from the $d_{xy}$ orbital and an almost identical but weaker contribution
at $(\pm 1,0)$ and $(0,\pm 1)$. (e) The susceptibility integrated around these momentum transfer vectors, shows the same trend at low energies. 
(d,f) In contrast, the orbital selective model yields spin fluctuations at low energies which are dominated by peaks at $(\pm 1,0)$ in the low energy range shown.
The enhancement of the spin fluctuations at $(\pm 1,0)$ in the superconducting state is clearly seen when plotted (f) as function of energy and (g) as an intensity map in momentum-energy space. (h) The spin fluctuations at $(1,0)$ are dominated by the  contributions of the $d_{yz}$ orbital. \label{fig_rpa_calc}}
\end{figure}

To confirm that spin fluctuations in FeSe for energies below 5 meV
have a $C_4$ component
as suggested in Figs. 3(a) and 3(b) and determine the impact of superconductivity (supplementary Figs. S7 and S8),
we carried out constant-energy and constant-wave-vector scans at $(1,0)$ and $(0,1)$ using a cold neutron
triple axis spectrometer (see Methods). Figures 4(a) and 4(b) show temperature difference plot below ($T=2$ K) and above ($T=12$ K) $T_c$ as a function of energy at $(1,0)$ and $(0,1)$, respectively.
In previous work on twinned samples,
superconductivity is found to induce a neutron spin resonance 
appearing below $T_c$ at $(1,0)$ and $(0,1)$ 
around $E\approx 3.6$ meV [Fig. 1(c)] \cite{wang16,MWMa17}. 
While Figure 4(a) shows clear evidence for the resonance at $E\approx 3.6$ meV with intensity 
reduction (negative scattering) below the mode 
indicating opening of a spin gap \cite{wang16,MWMa17}, an identical
temperature difference plot at $(0,1)$ in Fig. 4(b) yields no observable temperature
difference across $T_c$, and therefore no superconductivity-induced resonance and spin gap.
Figures 4(c) and 4(d) show wave vector scans along the $[H+1,0]$ and $[0,K+1]$ directions, respectively,
at $E=3.6$ meV. In the normal state ($T=12$ K), we see well defined peaks centered at
$(1,0)$ and $(0,1)$, consistent with Figs. 3(a) and 3(b). On cooling below $T_c$,
the scattering at $(1,0)$ increases in intensity and forms a resonance [Fig. 4(c)], while it
does not change across $T_c$ at $(0,1)$ [Fig. 4(d)].  Figures 4(e) and 4(f) show the same data
after correcting for background scattering and detwinning ratio $\eta$.  Similar to Figs. 4(c) and 4(d), we again
find that superconductivity induces a $C_2$ symmetric resonance on a background of approximately $C_4$ symmetric normal state magnetic scattering (supplementary Figs. S9). Thus, it is the highly anisotropic pairing state of FeSe that drives the $C_2$ symmetric magnetic scattering at these energies below $T_c$. Figures 1(f) and 1(g) summarize the key results of our INS experiments on detwinned FeSe.
The deviation of magnetic scattering intensity ratio at $(1,0)$ 
and $(0,1)$ from $3:1$ provides convincing evidence for the existence of an unexpected mode.
In the normal state, spin fluctuations have approximate $C_4$ symmetry near the 
resonance energy but become
$C_2$ symmetric for energies above 6 meV.  Upon entering the superconducting state,
a resonance with $C_2$ symmetry is formed at ${\bf Q}_{\rm AF}$ [Fig. 1(f)] (supplementary Figs. S10).

In order to achieve a theoretical understanding of the experimental results presented above, 
we start from an itinerant five-band model that quantitatively matches the low-energy electronic structure of FeSe in its nematic state \cite{bohmer,Coldea18,sprau17,kreisel17} and compute the magnetic scattering $S({\bf Q},E)\propto \chi^{\prime\prime}({\bf Q},E)/(1-e^{-E/k_BT})$ within a standard random phase approximation formulation \cite{kreisel17,Kreisel18,Kreisel15,Mukherjee15}. The spectral function at the Fermi level is presented in Fig. \ref{fig_rpa_calc}(a). As illustrated in Figs. \ref{fig_rpa_calc}(c) and \ref{fig_rpa_calc}(e), this ``plain vanilla'' approach completely fails as is evident, e.g., from the presence of scattering close to $(1,1)$, and a negligible $(1,0)$ - $(0,1)$ anisotropy. The latter properties can be traced to an improper balance of the three most important scattering channels [see Fig. \ref{fig_rpa_calc}(a)]. However, electronic interactions and associated self-energy effects are known to be important in FeSe, constituting an example of a Hund's metal \cite{Kostin2018}. Important properties of Hund's metals include the existence of orbital dependent mass renormalizations \cite{Georges_review,Medici_review,Roekeghem_review}, and an associated redistribution of the relative importance of different orbital dependent scattering channels in the spin susceptibility \cite{Ishizuka18}.

A simple means to incorporate the important effects of such orbital selectivity is through the introduction of orbital-dependent quasiparticle weights $Z_{\ell}<1$\cite{kreisel17,Kreisel18} leading to a modified bare susceptibility $\tilde \chi_{\ell_1 \ell_2 \ell_3 \ell_4}^0 ({\bf Q},E)$ given by 
\begin{equation}
 \tilde \chi_{\ell_1 \ell_2 \ell_3 \ell_4}^0 ({\bf Q},E)=\sqrt{Z_{\ell_1}Z_{\ell_2}Z_{\ell_3}Z_{\ell_4}}\;\chi_{\ell_1 \ell_2 \ell_3 \ell_4}^0 ({\bf Q},E).
 \label{eq_susc}
\end{equation}
In agreement with theoretical expectations \cite{Georges_review,Medici_review,Roekeghem_review,rong2018}, and earlier detailed studies of tunneling spectroscopy \cite{sprau17,Kostin2018}, we apply the hierarchy $Z_{xy}<Z_{xz}<Z_{yz}$, which shifts the relative importance of the dominant scattering vectors, as illustrated in Fig. \ref{fig_rpa_calc}(b), and thereby modifies the magnetic scattering. As seen in Fig. \ref{fig_rpa_calc}(d), the $d_{xy}$-dominated $(1,1)$-scattering is strongly reduced (because $Z_{xy}$ is the smallest), and the degree of $C_4$-symmetry breaking as seen by the difference in the scattering intensities at $(\pm 1,0)$ versus $(0,\pm 1)$ is strongly enhanced (because $Z_{xz}<Z_{yz}$), as seen explicitly by the dashed lines in 
Fig. \ref{fig_rpa_calc}(f) (Supplementary Fig. S1).

In the superconducting state, we employ the gap structure identical to the one of Refs. \cite{sprau17,kreisel17} which is known to faithfully describe the gap in FeSe, and modify the bare susceptibility accordingly \cite{Kreisel15,kreisel17}. When entering the highly anisotropic superconducting state, generated by the orbital-selective spin fluctuations \cite{kreisel17,sprau17}, a neutron resonance is exhibited solely at the $(\pm 1,0)$ position as seen from Figs.~\ref{fig_rpa_calc}(f) and \ref{fig_rpa_calc}(g), in agreement with  experiments. The associated neutron resonance is highly orbital-selective with predominant $d_{yz}$ character, as seen by the orbital-resolved spin susceptibilities plotted in Fig.~\ref{fig_rpa_calc}(h). Therefore, both the very strong $C_4$-symmetry breaking in the 5-10 meV range and the unidirectional neutron resonance observed experimentally are captured by the itinerant orbital-selective scenario. 

This approach, however, does not provide an explanation of the emergence of the localized approximately $C_4$-symmetric spin excitations near $E=3$ meV as shown in Figs. \ref{Fig1}(f), \ref{Fig1}(g) and \ref{Fig3}(a), \ref{Fig3}(b). There are several possible scenarios for this remarkable discovery. First, it is possible that self-energy effects in FeSe have a significantly more complicated functional form that cannot be simply captured by including energy- and momentum-independent $Z$-factors. Second, there is a possibility of impurity-generated low-energy spectral weight similar to the case of cuprates where vortices and disorder have been shown to generate localized modes in a restricted low-energy regime \cite{lake2001,kimura2003,andersen10}.
A counter-argument to disorder-based scenario, however, is the high quality of the FeSe crystals used in the current experiment (Supplementary Fig. S2).

Finally, if part of the spin excitations in FeSe arise from a local moment quantum paramagnet \cite{FWang15,Glasbrenner15,She17}, the $C_2$ symmetric AF collinear order competes with 
the $C_4$ symmetric N\'{e}el order across the nematic ordering temperature $T_s$ \cite{Qwang16}. 
In this picture, the $C_4$ symmetric low-energy magnetic excitations with spin-wave ring-like features
in detwinned FeSe may simply be the remnant of the localized moment not directly associated with 
Fermi surface nesting and itinerant electrons. 

Regardless of the microscopic origin of the $C_4$ spin excitations, our data support the notion that the spin fluctuations in the nematic phase of FeSe are, generally, highly anisotropic, and is consistent with superconductivity being driven by spin fluctuations arising mainly from the $d_{yz}$ orbital states. Our measurements highlight the need for a quantitative understanding of both the extreme spin anisotropy, as well as the emergence of $C_4$-symmetric magnetic excitations at the very lowest energies. Progress in this direction may well shed new light on the role of electronic correlations in FeSe, in particular, and the origin of unconventional superconductivity in interacting systems, in general.

\noindent {\bf Online content}

Any methods, additional references, Nature Research reporting
summaries, source data, statements of data availability and associated accession codes are available at:

\noindent {\bf Acknowledgements}

We thank D.  Abernathy, Q. Wang, Y. Hao and H. Hu for useful discussions. 
The neutron-scattering work at Rice University was supported by the U.S. 
DOE, BES DE-SC0012311 (P.D.). The single-crystal
synthesis work was supported by the Robert
A. Welch Foundation Grant No. C-1839 (P.D.). 
C. B. and Y. C. are supported by the Department of Energy 
Grant No. DE-FG02-08ER46544.
B.~M.~A. acknowledges financial support from the Carlsberg Foundation.
P. J. H.  was supported by the Department of Energy under Grant No. DE-FG02-05ER46236.
This research used resources at the High Flux Isotope Reactor and Spallation Neutron Source, a DOE Office of Science User Facility operated by the Oak Ridge National Laboratory. Access to MACS was provided by the Center for High Resolution Neutron Scattering, a partnership between the National Institute of Standards and Technology and the National Science Foundation under Agreement No. DMR-1508249.

\noindent {\bf Author contributions}

\noindent {X.Y.L., T.C., and P.D. conceived the project. T.C. prepared all the FeSe single crystal samples. BaFe$_2$As$_2$ single crystals are prepared by T.C., X.Y.L., R.Z., Y.L., Y.R., Y.W. Neutron scattering experiments on twinned samples are carried out by
T.C., Y.C., Y.M.Q., C.B., and P.D. at NCNR. 
Neutron scattering experiments on detwinned samples
are carried out by T.C., J.P., T.G.P., J.R.S., H.B.C., and P.D. at ORNL, ISIS, 
and MLZ. Theoretical analysis was performed by A.G., B.M.A., and P.J.H. The entire project was supervised by P.D.  The manuscript is written by P.D., T.C., A.K., B.M.A., and P.J.H. All authors made comments.  The authors declare no competing financial interests.}  

\noindent {\bf Additional information}

\noindent {Supplementary information is available for this paper at.}

\noindent {\bf Reprints and permissions information} is available at www.nature.com/reprints.

\noindent {\bf Correspondence and requests for materials} should be addressed to X.Y.L. or P.D.

\noindent {\bf Publisher’s note:} Springer Nature remains neutral with regard to jurisdictional claims in published maps and institutional affiliations.

\noindent {\bf Methods}

\noindent{{\bf Experimental Setups} 
Elastic neutron experiments were carried out on the HB-3A four-Circle diffractometer at the High-Flux-Isotope Reactor (HFIR), Oak Ridge National Laboratory (ORNL), United States to first check if the method works well in detwinning FeSe on a single piece of BaFe$_2$As$_2$ (Supplementary Fig. S3). HB-3A uses a silicon monochromator and a scintillator-based 2D Anger Camera. We define $(H,K,L)=(q_xa/2{\pi},q_yb/2{\pi},q_zc/2{\pi})$ in reciprocal lattice units (r.l.u.) using the orthorhombic lattice notation for FeSe, where $a\approx 5.33$ \AA, $b\approx 5.31$ \AA, $c=5.486$ \AA\ \cite{bohmer}.}

Our INS experiments on twinned samples were done on MACS cold triple axis spectrometer at NIST center for neutron scattering at Gaithersburg, Maryland. MACS spectrometer has a double focusing pyrolytic graphite 
[PG(002)] monochoromator 
and multi detectors. We used $E_f = 3.7$ meV 
with a BeO filter after the sample and a Be filter before the monochromator for energy transfers below $E=1.5$ meV.

Our INS experiments on detwinned samples
were carried out on the PANDA cold neutron and PUMA thermal neutron triple-axis spectrometers, at MLZ, Garching, Germany \cite{Lu14}, and on 
the MAPS time-of-flight chopper spectrometer, at ISIS, Rutherford-Appleton laboratory, Didcot, United Kingdom \cite{Lu18}. 

For PANDA experiments, a
double-focused pyrolytic graphite [PG(002)] monochromator and analyzer with fixed scattered neutron energy $E_{f} = 5.1$ meV were used with collimations of none-40$^{\prime}$-40$^{\prime}$-none for inelastic
measurements.  For elastic measurements, we used $E_{f} = 4.39$ meV with collimations of
80$^{\prime}$-80$^{\prime}$-80$^{\prime}$-80$^{\prime}$.
For thermal neutron measurements on PUMA, we used $E_{f} = 14.69$ meV with double focusing monochromator and analyzer and no collimators. 
 For MAPS neutron time-of-flight measurements, we  used an incident beam energy of $E_i=38$ meV with the incident beam along the $c$-axis of the crystal.

{\noindent {\bf Sample growth and preparation} The high quality FeSe single crystals used in the experiments are grown by a chemical vapor transport method. Fe and Se powder are sealed in quartz tubes with KCl-AlCl$_3$ flux. The growth takes 28 days in a temperature gradient from 330$^\circ$C to 400$^\circ$C. Typical samples are \mbox{$1\times 1$ mm$^2$} in area and $<0.1$ mm in thickness. 
The square-shaped BaFe$_2$As$_2$ crystals are grown using flux method \cite{Lu18}.
They are aligned using a Laue camera and cut along the tetragonal $[1,1,0]$
and $[1,-1,0]$ directions by a high-precision wire saw. 
Since single crystals of FeSe have one natural edge 45$^\circ$ rotated from the orthorhombic $a_o$ direction, we can use an optical method to co-align FeSe on the surface of BaFe$_2$As$_2$. Given our intent to measure spin excitations in detwinned FeSe, we aligned and glued (with CYTOP type-M) about 300 small pieces FeSe single crystals on many pieces of big BaFe$_2$As$_2$ single crystals (Supplementary Figs. S1-S3).}

\noindent {\bf Data availability}

\noindent {The data that support the plots in this paper and other findings of this study are
available from the corresponding authors upon reasonable request.}

\end{document}